\documentclass[journal,twocolumn,10pt]{IEEEtran}
\usepackage[T1]{fontenc}
\usepackage{cite}
\usepackage{amssymb,amsfonts}
\usepackage{algorithmic}
\usepackage{textcomp}
\usepackage[usenames, dvipsnames]{color}
\usepackage{subcaption}
\usepackage{graphicx}  

\ifCLASSINFOpdf
\else
\fi
\usepackage{amsmath}
\begin{document}
%
\title{Linear Receivers for Massive MIMO Systems with One-Bit ADCs}
%
%
%

\author{Ly~V.~Nguyen
	 and Duy~H.~N.~Nguyen, \emph{Senior Member, IEEE}
	\thanks{The authors are with the Department of Electrical and Computer Engineering and the Computational Science Research Center, San Diego State University, San Diego, CA, USA 92182. E-mail: vnguyen6@sdsu.edu, duy.nguyen@sdsu.edu.}
}

\maketitle

\begin{abstract}
	In this letter, we propose three linear receivers including Bussgang-based Maximal Ratio Combining (BMRC), Bussgang-based Zero-Forcing (BZF), and Bussgang-based Minimum Mean Squared Error (BMMSE) for massive MIMO systems with one-bit analog-to-digital converters (ADCs). Closed-form expressions of the proposed receivers are obtained by using the Bussgang decomposition to cope with the non-linear effect of the one-bit ADCs. Simulation results show significantly lower bit error rate floors obtained by the proposed receivers than those of conventional linear receivers.
\end{abstract}


%
\IEEEpeerreviewmaketitle

\section{Introduction}
Massive multiple-input multiple-output (MIMO) technology is being considered to be one of the disruptive solutions for 5G networks \cite{Boccardi2014Five,andrews2014will}. This is due to the capability of massive MIMO systems in boosting the throughput and energy efficiency by several magnitudes of enhancement over conventional MIMO systems \cite{ngo2013energy,Hoydis2013massive}. However, a massive MIMO system requires a large number of radio-frequency (RF) chains, which significantly increases the power consumption and hardware complexity \cite{Yang2013total}. Among the components of an RF chain, the analog-to-digital converter (ADC) is a power-hungry device since its power consumption increases exponentially with the number of bits per sample and linearly with the sampling rate \cite{walden1999analog,Murmann}. In millimeter-wave MIMO systems, the power consumption issue is even more serious due to the high sampling rate required for large bandwidths.

A promising solution for reducing the power consumption and hardware complexity is to use low-resolution ADCs, e.g, $1$-bit ADCs. The structure of a $1$-bit ADC is as simple as a single comparator, which means automatic gain controls are not required. Several other RF components such as mixers, frequency synthesizers and local oscillators can also be eliminated in some system architectures \cite{Donnell2005ultraWideband,Mezghani0217mmWave}. Hence, the use of $1$-bit ADCs can significantly reduce both the power consumption and hardware complexity.

Unfortunately, the non-linear effect of low-resolution ADCs makes the channel estimation and data detection tasks much more challenging. Numerous nonlinear detection methods have been proposed in the literature, e.g., \cite{choi2016near,Xiong2017Low,wen2016bayes,Jeon2018One}, to deal with such nonlinearities. The authors in \cite{choi2016near} proposed a two-stage near maximum-likelihood (ML) estimator, where the ML detection problem is relaxed into a convex optimization program in the first stage and then the second stage reduces the number of candidate transmit vectors based on the output of the first stage. The Generalized Approximate Message Passing (GAMP) algorithm and sphere decoding technique were employed in \cite{Xiong2017Low} and \cite{Jeon2018One}, respectively. A joint channel-and-data estimation method was proposed in \cite{wen2016bayes} by using Bayesian inference and the GAMP algorithm. These nonlinear detection methods were shown to be effective, but they suffer from high computational complexity. In systems where low computational complexity and fast detection are required, linear receivers are often preferable since the multiuser detection task is accomplished by simply multiplying the received signal vector with a combining matrix.

In this letter, we study linear receivers for massive MIMO systems with $1$-bit ADCs. All of the existing works take one of the following two strategies: (i) ignore the effect of $1$-bit ADCs and use the linear receiver structures of systems with infinite-resolution ADCs, e.g.,\cite{choi2016near,li2017channel,Jacobsson2017Throughput}; or (ii) using an approximate model for the $1$-bit ADC to construct other linear receiver structures, e.g., \cite{Liu2018Asymptotic,mezghani2007modified}. Here, we exploits the Bussgang decomposition  \cite{bussgang1952crosscorrelation} to propose three new simple, yet efficient linear receivers. A major advantage of the Bussgang decomposition is that it can provide an exact linear input-output relationship of the $1$-bit ADC massive MIMO system. Simulation results show that our proposed linear receivers based on the Bussgang decomposition significantly outperform existing ones. The rest of this letter is organized as follows. Section \ref{sec_II} introduces the system model. In Section \ref{sec_III}, we first present several conventional linear receivers in literature and then use the Bussgang decomposition to propose three new linear receivers. Numerical results are provided in Section \ref{sec_IV} to show the superiority of the proposed linear receivers over conventional ones. Finally, Section \ref{sec_VI} concludes the paper.

\textit{Notation}: Upper-case and lower-case boldface letters denote matrices and column vectors, respectively. $\mathbb{E}[\cdot]$ represents expectation. The transpose and conjugate transpose are denoted by $[\cdot]^T$ and $[\cdot]^H$, respectively. The notation $\Re\{.\}$ and $\Im\{.\}$ respectively denotes the real and imaginary parts of the complex argument.
\section{System Model}
\label{sec_II}
We consider an uplink massive MIMO system with $K$ single-antenna users and an $N$-antenna base station, where it is assumed that $N \geq K$. Each receive antenna at the base station is equipped with a pair of $1$-bit ADCs for the in-phase and quadrature signal components. Let $\mathbf{x} = [x_1, \ldots,x_K]^T \in \mathbb{C}^K$ be the transmitted signal vector, where $x_k$ is the symbol transmitted from the $k^{\text{th}}$ user, $k = 1, \ldots, K$. Each symbol $x_k$ is drawn from a constellation $\mathcal{M}$ under the power constraint $\mathbb{E}[|x_k|^2]=1$. Let $\mathbf{h}_k \in \mathbb{C}^N$ be the uplink channel from user $k$ and $\mathbf{H} = [\mathbf{h}_1,\mathbf{h}_2,\ldots,\mathbf{h}_K] \in \mathbb{C}^{N\times K}$, which is assumed to be block-fading and known at the base station. An efficient channel estimator based on the Bussgang decomposition for massive MIMO systems with $1$-bit ADCs can be found in \cite{li2017channel}. We assume the Rayleigh fading channel with independent and identically distributed (i.i.d.) elements, and each channel element is distributed as $\mathcal{CN}(0,1)$.
Let $\mathbf{r} = [r_1,\ldots,r_{N}]^T \in \mathbb{C}^N$ be the analog received signal vector at the base station, which is given as 
\begin{equation}
\mathbf{r} = \mathbf{H}\mathbf{x}+\mathbf{z},\\
\end{equation}
where $\mathbf{z} = [z_1,\ldots,z_{N}]^T \in \mathbb{C}^{N}$ is the noise vector. The noise elements are assumed to be i.i.d. with $z_i \sim \mathcal{CN}(0,N_0)$. Let $\mathbf{y} = [y_1,\ldots,y_{N}]^T$ be the digital received signal vector and let the $1$-bit ADC be represented by the $\operatorname{sign}(.)$ operator where $\operatorname{sign}(a) = 1$ if $a\geq 0$, and $\operatorname{sign}(a) = -1$ if $a <0$, then $y_i = \operatorname{sign}(\Re\{r_i\}) + j\operatorname{sign}(\Im\{r_i\})$, $i=1,\ldots,N$. The signal-to-noise ratio (SNR) is defined as $\rho = 1/N_0$.
\section{Linear Receivers for $1$-bit ADCs}
\label{sec_III}
This section aims to examine different types of linear receivers for massive MIMO systems with $1$-bit ADCs. We first present conventional linear receivers and then use the Bussgang decomposition to propose three new ones including Bussgang-based Maximal Ratio Combining (BMRC), Bussgang-based  Zero-Forcing (BZF), and Bussgang-based  Minimum Mean Squared Error (BMMSE). 

Given a received signal vector $\mathbf{y}$ and a linear receiver represented by a combining matrix $\mathbf{W} = [\mathbf{w}_1,\mathbf{w}_2,\ldots,\mathbf{w}_K]^T$, the demultiplexing task is performed as
\begin{equation}
\tilde{\mathbf{x}} = [\tilde{x}_1,\tilde{x}_2,\ldots,\tilde{x}_K]^T = \mathbf{W}\mathbf{y}.
\label{eq_demultiplexing}
\end{equation}
The signal $\tilde{\mathbf{x}}$ is then equalized before symbol-by-symbol detection is performed. In the following, we present different structures of the combining matrix $\mathbf{W}$.
\subsection{Conventional Linear Receivers}
A straightforward strategy to obtain linear receivers for massive MIMO systems with $1$-bit ADCs is to simply ignore the non-linear effect of the $1$-bit ADCs and use the conventional linear receivers of massive MIMO systems with infinite-resolution ADCs as follows:
\begin{itemize}
	\item Maximal Ratio Combining (MRC) receiver
	\begin{equation}
	\mathbf{W}_{\mathtt{MRC}} = \mathbf{H}^H.
	\end{equation}
	\item Zero-Forcing (ZF) receiver
	\begin{equation}
	\mathbf{W}_{\mathtt{ZF}} = \big(\mathbf{H}^H\mathbf{H}\big)^{-1}\mathbf{H}^H.
	\end{equation}
	\item Minimum Mean Squared Error (MMSE) receiver
	\begin{equation}
	\mathbf{W}_{\mathtt{MMSE}} = \big(\mathbf{H}^H\mathbf{H} + N_0\mathbf{I}_K\big)^{-1}\mathbf{H}^H.
	\end{equation}
\end{itemize}

In another strategy, the nonlinear effect of the $1$-bit ADCs can be linearized by the Additive Quantization Noise Model (AQNM) \cite{Fletcher2007roubust,Orhan2015low} as
\begin{equation}
\mathbf{y} = \kappa\mathbf{r} + \mathbf{q} = \kappa \mathbf{H}\mathbf{x} + \kappa \mathbf{z} + \mathbf{q},
\label{eq_AQNM_model}
\end{equation}
where $\kappa = 1 - \alpha$ and $\alpha$ is the inverse of the signal-to quantization-noise ratio, which is approximated as $\alpha \approx 0.3634$ for $1$-bit ADCs \cite{Orhan2015low}. The quantization distortion $\mathbf{q}$ is  uncorrelated to $\mathbf{r}$ and treated as an additive Gaussian noise with $\mathbf{q} \sim \mathcal{CN}(\mathbf{0},\boldsymbol{\Sigma}_q)$ where
$
\boldsymbol{\Sigma}_q = \alpha \kappa \operatorname{diag}(\mathbf{H}\mathbf{H}^H+N_0\mathbf{I}_N).
$
The MMSE receiver for the model (\ref{eq_AQNM_model}) is given as \cite{Liu2018Asymptotic}
\begin{equation}
\mathbf{W}_{\mathtt{AQNM-MMSE}} = \mathbf{H}^H\bigg(\mathbf{H}\mathbf{H}^H + \frac{1}{\kappa^2}\boldsymbol{\Sigma}_q + N_0\mathbf{I}_N\bigg)^{-1}.
\end{equation}

Another approximate MMSE receiver for quantized MIMO systems, which is referred to as  ``Wiener Filter on Quantized
data'' (WFQ), is proposed in \cite{mezghani2007modified} as
\begin{equation}
\mathbf{W}_{\mathtt{WFQ}} = \mathbf{H}^H\Big(\kappa \mathbf{\Sigma}_{r} + \alpha \operatorname{diag}(\mathbf{\Sigma}_r)\Big)^{-1},
\end{equation}
where $\mathbf{\Sigma}_{r} = \mathbf{H}\mathbf{H}^H + N_0\mathbf{I}_N$ is the covariance matrix of $\mathbf{r}$.

Once a combining matrix $\mathbf{W}$ has been derived, the demultiplexing task can be performed as in (\ref{eq_demultiplexing}). Then, the signal $\tilde{\mathbf{x}}$ is equalized as
\begin{equation}
\check{x}_k = \frac{\tilde{x}_k}{\mathbf{w}_k^T\mathbf{h}_k},
\end{equation}
where $\mathbf{w}_k$ is the $k^{\text{th}}$ column of the corresponding combining matrix, which can be $\mathbf{W}_{\mathtt{MRC}}$, $\mathbf{W}_{\mathtt{MMSE}}$, $\mathbf{W}_{\mathtt{AQNM-MMSE}}$, or $\mathbf{W}_{\mathtt{WFQ}}$. Note that if the ZF combining matrix $\mathbf{W}_{\mathtt{ZF}}$ is used, this equalization step is not necessary because the ZF receiver cancels out interference. Since the norm square of $\check{\mathbf{x}} = [\check{x}_1,\check{x}_2,\ldots,\check{x}_K]^T$ may not equal $K$, the signal $\check{\mathbf{x}}$ should be rescaled as \cite{choi2016near}
\begin{equation}
	\acute{\mathbf{x}}= [\acute{x}_1, \acute{x}_2,\ldots,\acute{x}_K]^T = \sqrt{K}\frac{\check{\mathbf{x}}}{\|\check{\mathbf{x}}\|_2}.
	\label{eq_rescaled_signal}
\end{equation}
The above rescaling step is not necessary when PSK modulation is used, but it matters for QAM modulation. Finally, the signal $\acute{\mathbf{x}}$ can be used for symbol-by-symbol detection as
\begin{equation}
	\hat{x}_k = \arg\max_{x\in\mathcal{M}} |\acute{x}_k - x|^2.
	\label{eq_sym_by_sym_detection}
\end{equation}
\subsection{Proposed Bussgang-based Linear Receivers}
Here, we exploit the Bussgang decomposition to linearize the system model $\mathbf{y} = \operatorname{sign}(\mathbf{r})$ and then use the linearized model to propose new MRC, ZF, and MMSE receiver structures. Following the Bussgang decomposition, the system model $\mathbf{y} = \operatorname{sign}(\mathbf{r})$ can be rewritten as $\mathbf{y} = \mathbf{F}\mathbf{r} + \mathbf{e}$ \cite{mezghani2012capacity} where $\mathbf{e}$ is the quantization distortion, which is uncorrelated to $\mathbf{r}$, i.e., $\mathbb{E}\big[\mathbf{r}\mathbf{e}^H\big] = \mathbb{E}\big[\mathbf{r}\big]\mathbb{E}\big[\mathbf{e}^H\big]$, and 
\begin{equation}
\mathbf{F} = \sqrt{\frac{2}{\pi}}\operatorname{diag}(\boldsymbol{\Sigma}_{r})^{-\frac{1}{2}}.
\end{equation} 
Let $\mathbf{A} = \mathbf{F}\mathbf{H}$ and $\mathbf{n} = \mathbf{F}\mathbf{z} + \mathbf{e}$, the system model becomes
\begin{equation}
\mathbf{y} = \mathbf{A}\mathbf{x} + \mathbf{n},
\label{eq_Bussgang_based_model}
\end{equation}
where $\mathbf{A} = \sqrt{2/\pi}\operatorname{diag}(\boldsymbol{\Sigma}_{r})^{-\frac{1}{2}}\mathbf{H}$ is the effective channel and $\mathbf{n}$ is the effective noise, which is modeled as Gaussian \cite{mezghani2012capacity} with zero mean and covariance matrix
\begin{equation}
\begin{split}
\boldsymbol{\Sigma}_{n} = &\frac{2}{\pi}\Big[\operatorname{arcsin}\Big(\operatorname{diag}(\mathbf{\Sigma}_{r})^{-\frac{1}{2}}\mathbf{\Sigma}_{r}\operatorname{diag}(\mathbf{\Sigma}_{r})^{-\frac{1}{2}}\Big)-\\
&\quad\;\;\operatorname{diag}(\mathbf{\Sigma}_{r})^{-\frac{1}{2}}\mathbf{\Sigma}_{r}\operatorname{diag}(\mathbf{\Sigma}_{r})^{-\frac{1}{2}}+\\
&\quad\;\;N_0\operatorname{diag}(\mathbf{\Sigma}_{r})^{-1}\Big].
\end{split}
\end{equation}
Note that $\operatorname{arcsin}(\mathbf{C}) = \operatorname{arcsin}(\Re\{\mathbf{C}\}) + j\operatorname{arcsin}(\Im \{\mathbf{C}\})$ for any complex matrix $\mathbf{C}$, and the operation $\operatorname{arcsin}(\mathbf{.})$ of a real matrix is applied separately on each element of that matrix. 

By utilizing the effective channel $\mathbf{A}$, a Bussgang-based MRC (BMRC) receiver and a Bussgang-based ZF (BZF) can be obtained as
\begin{equation}
\mathbf{W}_{\mathtt{BMRC}} = \mathbf{A}^H = \sqrt{\frac{2}{\pi}}\mathbf{H}^H\operatorname{diag}(\boldsymbol{\Sigma}_r)^{-\frac{1}{2}},
\end{equation}
and
\begin{equation}
\begin{split}
\mathbf{W}_{\mathtt{BZF}} &= (\mathbf{A}^H\mathbf{A})^{-1}\mathbf{A}^H\\
&=\sqrt{\frac{\pi}{2}} \Big(\mathbf{H}^H\operatorname{diag}(\boldsymbol{\Sigma}_r)^{-1}\mathbf{H}\Big)^{-1}\mathbf{H}^H\operatorname{diag}(\boldsymbol{\Sigma}_r)^{-\frac{1}{2}}.
\end{split}
\end{equation}

We now drerive the MMSE receiver for this Bussgang-based system model. The Bussgang-based MMSE (BMMSE) receiver can be obtained by solving the following optimization problem:
\begin{equation}
\begin{aligned}
& \underset{\{\mathbf{W}\}}{\operatorname{minimize}} 
& & \mathbb{E}\big[\|\mathbf{x} - \mathbf{W}\mathbf{y}\|_2^2\big].
\end{aligned}
\label{eq_BMMSE_problem}
\end{equation}
Let $f =\mathbb{E}\big[\|\mathbf{x} - \mathbf{W}\mathbf{y}\|_2^2\big]$ be the objective function to be minimized and $\mathbf{B} = \mathbf{I} - \mathbf{WA}$, so we have
\begin{align}
f &= \mathbb{E}\big[\|\mathbf{B}\mathbf{x} - \mathbf{W}\mathbf{n}\|_2^2\big]\\
&=\mathbb{E}\bigg[\operatorname{trace}\Big\{\mathbf{B}\mathbf{x}\mathbf{x}^H\mathbf{B}^H - \mathbf{B}\mathbf{x}\mathbf{n}^H\mathbf{W}^H -\notag \\
& \qquad \qquad \quad \mathbf{W}\mathbf{n}\mathbf{x}^H\mathbf{B}^H + \mathbf{W}\mathbf{n}\mathbf{n}^H\mathbf{W}^H\Big \}\bigg]\label{eq_18}\\
&=\operatorname{trace}\big\{\mathbf{B}\mathbf{B}^H\big\} + \operatorname{trace}\big\{ \mathbf{W}\boldsymbol{\Sigma}_n\mathbf{W}^H\} \label{eq_19}.
\end{align}
Note that (\ref{eq_18}) is equivalent to (\ref{eq_19}) because $\mathbb{E}\big[\mathbf{x}\mathbf{x}^H\big] = \mathbf{I}$, $\mathbb{E}\big[\mathbf{n}\mathbf{n}^H\big] = \boldsymbol{\Sigma}_n$, and $\mathbb{E}\big[\mathbf{x}\mathbf{n}^H\big] = \mathbf{0}$. We have $\mathbb{E}\big[\mathbf{x}\mathbf{n}^H\big] = \mathbf{0}$ because of
\begin{equation*}
	\mathbb{E}\big[\mathbf{x}\mathbf{n}^H\big] = \mathbb{E}\big[\mathbf{x}(\mathbf{F}\mathbf{z} + \mathbf{e})^H\big]
	= \mathbb{E}\big[\mathbf{x}\mathbf{z}^H\big]\mathbf{F} + \mathbb{E}\big[\mathbf{x}\mathbf{e}^H\big],
\end{equation*}
where $\mathbb{E}\big[\mathbf{x}\mathbf{z}^H\big] = \mathbb{E}\big[\mathbf{x}\big] \mathbb{E}\big[\mathbf{z}^H\big] = \mathbf{0}$, and $\mathbb{E}\big[\mathbf{x}\mathbf{e}^H\big] = \mathbf{0}$ since
\begin{equation*}
	\begin{cases}
	\mathbb{E}\big[\mathbf{r}\mathbf{e}^H\big] = \mathbf{H}\,\mathbb{E}\big[\mathbf{x}\mathbf{e}^H\big] + \mathbb{E}\big[\mathbf{z}\mathbf{e}^H\big],\\
	\mathbb{E}\big[\mathbf{r}\mathbf{e}^H\big] = \mathbb{E}\big[\mathbf{r}\big]\mathbb{E}\big[\mathbf{e}^H\big] = \mathbf{0},\\
	\mathbb{E}\big[\mathbf{z}\mathbf{e}^H\big] = \mathbf{0}.
	\end{cases}
\end{equation*}
A similar proof of $\mathbb{E}\big[\mathbf{x}\mathbf{n}^H\big] = \mathbf{0}$ can also be found in \cite{Balevi2019One}.

The derivative of $f$ with respect to $\mathbf{W}$ is
\begin{equation}
\frac{\partial f }{\partial \mathbf{W}} = -\mathbf{A} + \mathbf{A}\mathbf{A}^H\mathbf{W}^H + \boldsymbol{\Sigma}_n\mathbf{W}^H,
\end{equation}
which when being set to zero yields
\begin{align}
&\mathbf{W}_{\mathtt{BMMSE}} = \mathbf{A}^H(\mathbf{A}\mathbf{A}^H + \boldsymbol{\Sigma}_n)^{-1}.
\end{align}
It can be seen that the structure of the optimal BMMSE receiver is similar to that of the MMSE receiver, except that the BMMSE receiver imposes a new effective channel and a new effective noise covariance. These differences come as the result of applying the Bussgang decomposition to the nonlinear 1-bit ADC system model.

Since the effective channel is $\mathbf{A}$, the equalization step is now performed as
\begin{equation}
\check{x}_k = \frac{\tilde{x}_k}{\mathbf{w}_k^T\mathbf{a}_k},
\end{equation}
where $\mathbf{w}_k$ is the $k^{\text{th}}$ column of the used combining matrix, which is $\mathbf{W}_{\mathtt{BMRC}}$ or $\mathbf{W}_{\mathtt{BMMSE}}$; and $\mathbf{a}_k$ is the $k^{\text{th}}$ column of $\mathbf{A}$. The rescaling step and symbol-by-symbol detection are the same as in (\ref{eq_rescaled_signal}) and (\ref{eq_sym_by_sym_detection}).

\section{Numerical Results}
\label{sec_IV}
\begin{figure}[t!]
	\centering
	\begin{subfigure}[t]{0.5\textwidth}
		\centering
		\includegraphics[scale=0.63]{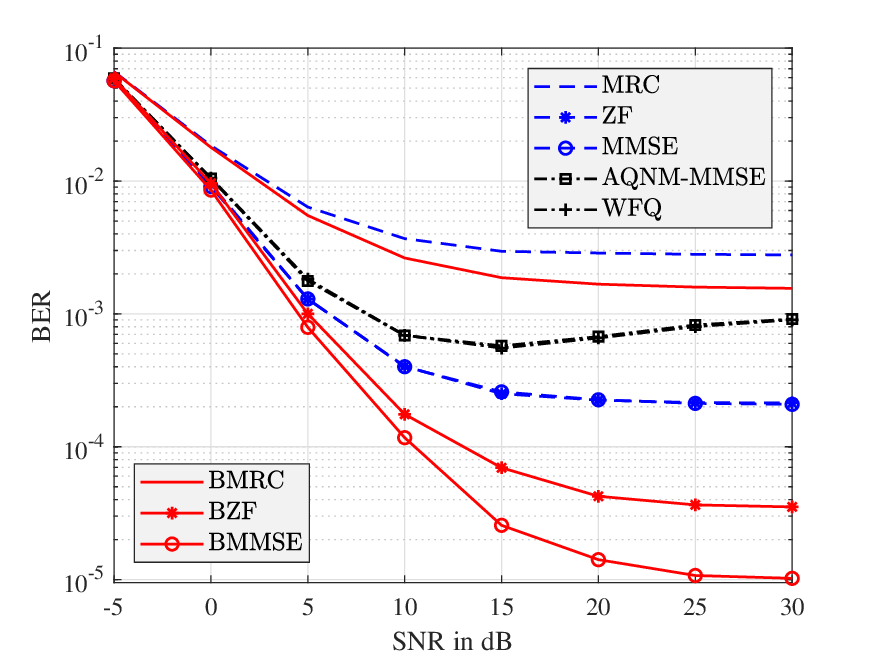}
		\caption{$K = 2$, $N = 16$, and QPSK modulation}
		\label{fig_1a}
	\end{subfigure}%

	\begin{subfigure}[t]{0.5\textwidth}
		\centering
		\includegraphics[scale=0.63]{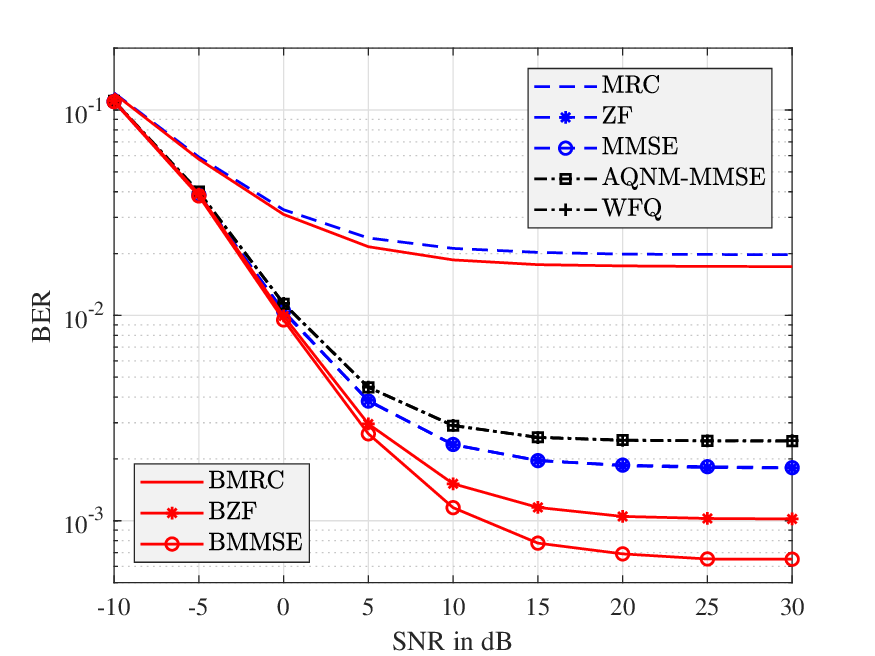}
		\caption{$K = 4$, $N = 64$, and $8$-PSK modulation}
		\label{fig_1b}
	\end{subfigure}
	\caption{Performance comparisons between the proposed and existing linear receivers with 1-bit ADCs. The proposed Bussgang-based schemes clearly show superior BER floors over the conventional counterparts with increasing SNR.}
	\label{fig_1}
\end{figure}
We use Monte Carlo simulations to numerically evaluate the performance of our proposed linear receivers. Fig. \ref{fig_1} presents the bit-error-rate (BER) comparison between the proposed and existing linear receivers with $K = 2$, $N = 16$, and QPSK modulation in Fig. \ref{fig_1a}; and $K = 4$, $N = 64$, and $8$-PSK modulation in Fig. \ref{fig_1b}. It can be seen that the proposed Bussgang-based linear receivers outperform their conventional counterparts. The BER floors of the proposed linear receivers are lower than those of conventional ones. These performance improvements are achieved thanks to the exact linear input-output relationship of massive MIMO systems with $1$-bit ADCs obtained by the Bussgang decomposition.

\begin{figure}
	\centering
	\includegraphics[scale=0.6]{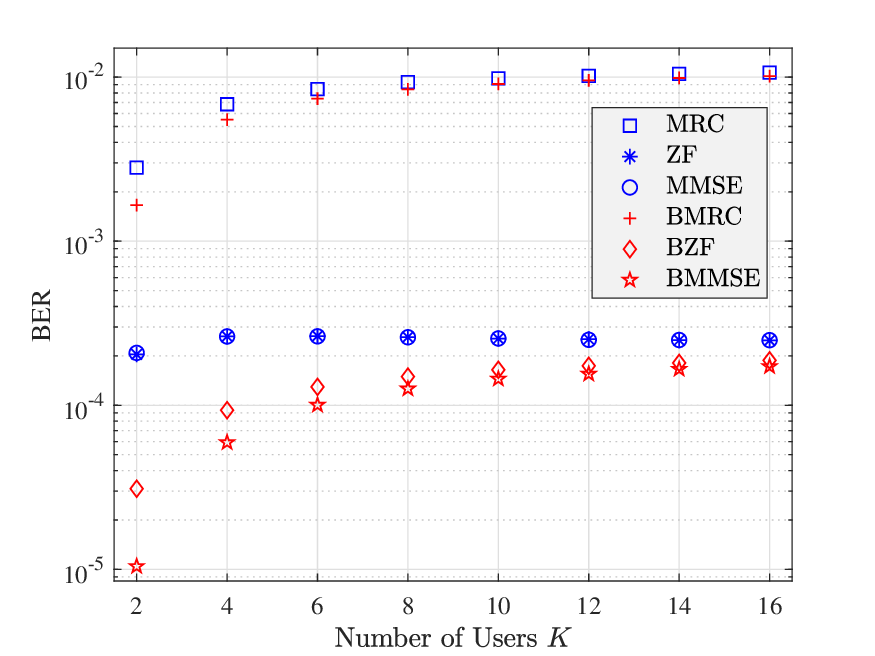}
	\caption{Error floor comparison for different numbers of users $K$ with QPSK signaling. The number of antennas at the base station is set to be $N = 8K$.}
	\label{fig_2}
\end{figure}

In Fig. \ref{fig_2}, we show the error floors for different numbers of users $K$. Each point in the figure is an error floor obtained at $30$ dB. We consider $K \in \{2,4,6,8,10,12,14,16\}$ and $N = 8K$. Here, we omitted the error floors of AQNM-MMSE and WFQ since they are higher than those of ZF and MMSE. It is observed that the proposed linear receivers always outperform their conventional counterparts, and the performance improvement is best seen when $K$ is not too large. As $K$ increases, the gap between the error floors tend to diminish, which can be explained as follows.

For large $K$, we have an approximation that $\mathbf{H}\mathbf{H}^H \approx K\mathbf{I}_N$, which yields $\boldsymbol{\Sigma}_r \approx (K+N_0)\mathbf{I}_N$, $\mathbf{A} \approx \sqrt{\gamma}\mathbf{H}$, and
\begin{equation}
\boldsymbol{\Sigma}_n \approx \big(1-\gamma K\big)\mathbf{I}_N,
\end{equation}
where $\gamma = 2/(\pi (K+N_0))$. These approximations lead to
\begin{align*}
\mathbf{W}_{\mathtt{BMRC}} &\approx \sqrt{\gamma}\mathbf{H}^H,\\
\mathbf{W}_{\mathtt{BZF}} & \approx \sqrt{\frac{1}{\gamma}}\Big(\mathbf{H}^H\mathbf{H}\Big)^{-1}\mathbf{H}^H,\\
\mathbf{W}_{\mathtt{BMMSE}} & \approx \sqrt{\frac{1}{\gamma}}\mathbf{H}^H\bigg(\mathbf{H}\mathbf{H}^H + \frac{1-\gamma K}{\gamma}\mathbf{I}_N\bigg)^{-1}.
\end{align*}
The above approximated Bussgang-based linear receivers are analogous to the conventional ones, albeit some scaling factor. This explains why their performances are approximately the same for large $K$.

\section{Conclusion}
\label{sec_VI}
In this letter, we have presented different conventional linear receivers for massive MIMO systems with $1$-bit ADCs. We have then proposed three new linear receiver structures by applying the Bussgang decomposition to linearizing the system model. Numerical results have shown that the proposed linear receivers outperform the existing ones much lower achievable BER floors.


%




\ifCLASSOPTIONcaptionsoff
  \newpage
\fi



%
\bibliographystyle{IEEEtran}
\bibliography{ref}

%








\end{document}